\documentclass[11pt,a4paper]{article}
\usepackage{jheppub}
\usepackage{setspace,braket}
\usepackage{amsmath, amssymb, bm}
\usepackage{etoolbox}

    \makeatletter
    \patchcmd{\maketitle}{\@fpheader}{}{}{}
\makeatother
\def\be{\begin{equation}}
\def\ee{\end{equation}}

\def\({\left(}
\def\){\right)}
\def\[{\left[}
\def\]{\right]}

\newcommand{\bea}{\begin{eqnarray}}
\newcommand{\eea}{\end{eqnarray}}

\def\d#1#2{\frac{\displaystyle #1}{\displaystyle #2}}
\def\no{\nonumber}

\numberwithin{equation}{section}
\renewcommand{\thesection}{\arabic{section}}

\begin{document}
\renewcommand{\thefootnote}{\fnsymbol{footnote}}

\title{Fractional phase transitions of RN-AdS black hole at Davies points}
\author[a,b]{Li-Hua Wang,}
\author[a]{Yun He,}
\author[a,b]{Meng-Sen Ma\footnote{Corresponding author. \\
		\indent\indent  E-mail address: mengsenma@gmail.com.}}
\affiliation[a]{Department of Physics, Shanxi Datong
University,  Datong 037009, China}
\affiliation[b]{Institute of Theoretical Physics, Shanxi Datong
University, Datong 037009, China}

\abstract{

We perform a study of phase transitions of RN-AdS black hole at its Davies points according to a generalized Ehrenfest classification of phase transition established on the basis of fractional derivatives. Davies points label the positions where heat capacity diverges. According to the usual Ehrenfest classification, second-order phase transitions occur there. For RN-AdS black hole, the Davies points can be classified into two types. The first type corresponds to the extreme values of temperature and the second type corresponds to the infection point(namely the critical point) of temperature. Employing the generalized Ehrenfest classification, we find that the orders of phase transition at the two types of Davies points are different. It is $3/2$-order for the first type and $4/3$-order for the second type. Thus this finer-grained classification can discriminate phase transitions that are supposed to be in the same category, which may provide some new insights toward a better understanding of black hole thermodynamics.

}
\maketitle
\onehalfspace

\renewcommand{\thefootnote}{\arabic{footnote}}
\setcounter{footnote}{0}
\section{Introduction}
\label{intro}

The works of Hawking and Bekenstein laid a firm foundation for the study of black hole thermodynamics. Phase transition and critical phenomena are important characteristics of usual thermodynamic systems. Therefore one may naturally consider the corresponding properties of black holes. Davies found that heat capacities of some black holes will diverge at certain points, according to the Ehrenfest classification which is a feature associated with a second-order phase transition\cite{Hut.379389.1977, Davies.13131355.1978, Sokolowski.11131120.1980}. These divergent points of heat capacities are called Davies points now. 

In general, the heat capacities of black holes can be expressed in the following form 
\be
C_X=\left.T\frac{\partial S}{\partial T}\right|_X=\left.T\frac{\partial S/\partial r_h}{\partial T/\partial r_h}\right|_X,
\ee
where $X$ represents a certain thermodynamic quantity and $r_h$ is the radius of the event horizon. Therefore, divergent heat capacity is equivalent to  $\partial T/\partial r_h=0$. In fact, according to this result, Davies points can be further classified into two types. For the first type, the Davies points correspond to extreme points of temperature, where temperatures may have maximal or minimal values. While for the second type, the Davies point lies at the infection point of temperature, which is also the critical points of black holes. 

Properties of various black holes near critical points have been studied extensively.  On the basis of AdS/CFT, phase structures of the charged AdS black holes were studied  in\cite{Chamblin.064018.1999, Chamblin.104026.1999}. Thermodynamic stability and phase transition of some charged black holes were analyzed by putting them in finite box \cite{Peca.124007.1999, Carlip.38273837.2003,Lu.133.2011}. By means of thermodynamic geometry, thermodynamic curvature can be calculated, from which one can ascertain the phase transition of black holes near the critical points\cite{Liu.054.2010, Banerjee.064024.2011, Cao.064015.2011, Wang.401404.2011, Niu.024017.2012, Wei.044014.2013}. Inspired by the idea in  \cite{Dolan.235017.2011} of treating the cosmological constant as pressure and its conjugate quantity as the thermodynamic volume, one can construct an extended phase space. In analogy to liquid-gas system, $P-V$ criticality and phase structures of many AdS black holes have been analyzed\cite{Kubiznak.033.2012, Cai.005.2013, Hendi.084045.2013, Chen.060401.2013, Mo.056.2013, Liu.179.2014, Ma.095001.2014, Xu.3074.2014, Ma.035024.2015, Xu.124033.2015, Wei.084015.2016, Ma.024052.2017, Miao.084051.2018, Xu.104022.2021}. In all these works, it concludes that the phase transition occurring at the critical point is of the second order. 

Hilfer proposed a kind of generalized Ehrenfest classification, which classifies phase transition according to the continuity of the fractional derivatives of free energy\cite{Hilfer.1992, Hilfer.2000}. Inspired by Hilfer's work, we reanalyzed the phase transition of RN-AdS black hole using this generalized Ehrenfest classification and found that the order of phase transition at the critical point should be of $4/3$, but not $2$ \cite{Ma.490495.2019}. At present, this generalized Ehrenfest classification is mainly used to deal with the phase transition at the critical(Davies) points, so they do not affect the original phase structure of black holes. Chabab and Iraoui  further studied AdS black holes in higher dimensions and without spherical symmetry and found that the order of fractional phase transition can be dependent on the geometry of spacetime\cite{Chabab.620430.2021,Chabab.2022}.

The two types of Davies points both correspond to divergent heat capacity. In other words, we cannot discriminate Davies points of the two types by means of heat capacity. Employing fractional derivatives, one can generalize the Ehrenfest classification to include phase transitions of fractional order. This finer-grained classification may reflect more richer phase structures of thermodynamic systems. 
We take the RN-AdS black hole as an example to analyze the phase structure at its Davies points. Considering that the temperatures of RN-AdS black hole have different behaviors at the two types of Davies points, we expect that there should be different phase structures. Therefore, it is our concern in this letter whether phase structures at the two types of Davies points can be distinguished by means of this generalized Ehrenfest classification. The second motivation is that we want to know whether the order of phase transition at the critical point is the same in extended phase space and the non-extended phase space.

The paper is arranged as follows. We first simply review RN-AdS black hole and give some thermodynamic quantities in section 1. In section 2 we analyze the phase structures at the first type of Davies point. It refers to two Davies points on the left-hand side and the right-hand side. In section 3 we discuss phase structures of the RN-AdS black hole at the critical point. At last, we summarize our results and discuss the possible future study.

\section{The RN-AdS black hole}

RN-AdS black hole is a static spherically symmetric spacetime, the line element of which is
\be
ds^2=-f(r)dt^2+f^{-1}(r)dr^2+r^2d\Omega^2.
\ee
The metric function is
\be
f(r)=1-\d{2M}{r}+\d{Q^2}{r^2}-\d{\Lambda r^2}{3},
\ee
where the parameters $M$,$Q$, and $\Lambda=-3/L^2$ are the black hole mass, electric charge and the cosmological constant, respectively.

According to $f(r_h)=0$, we can find the position of the event horizon. We do not consider the extended phase space in this paper.  There is no thermodynamic volume and pressure in the first law of thermodynamics, which is 
\be
dM=TdS+\phi dQ.
\ee
In this case, $M$ is understood as the internal energy. Through Legendre transformation, we can obtain the Helmholtz free energy $F=M-TS$, which satisfies the differential relation
\be
dF=-SdT+\phi dQ.
\ee
Thus, $F=F(T,Q)$ is a  thermodynamic characteristic function, from which we can derive all other thermodynamic quantities. We will consider the canonical ensemble and take $(T,Q)$ as the thermodynamic variables. 

The temperature of RN-AdS black hole is 
\be\label{TRNADS}
T=\frac{L^2 r_h^2+3 r_h^4-L^2 Q^2}{4 \pi  L^2 r_h^3}.
\ee

The entropy satisfies the area law, $S=A/4=\pi r_h^2$. And the electric potential at the horizon measured at infinity is $\phi=Q/r_h$.

\begin{figure}
	\centering{
		\includegraphics[width=5cm]{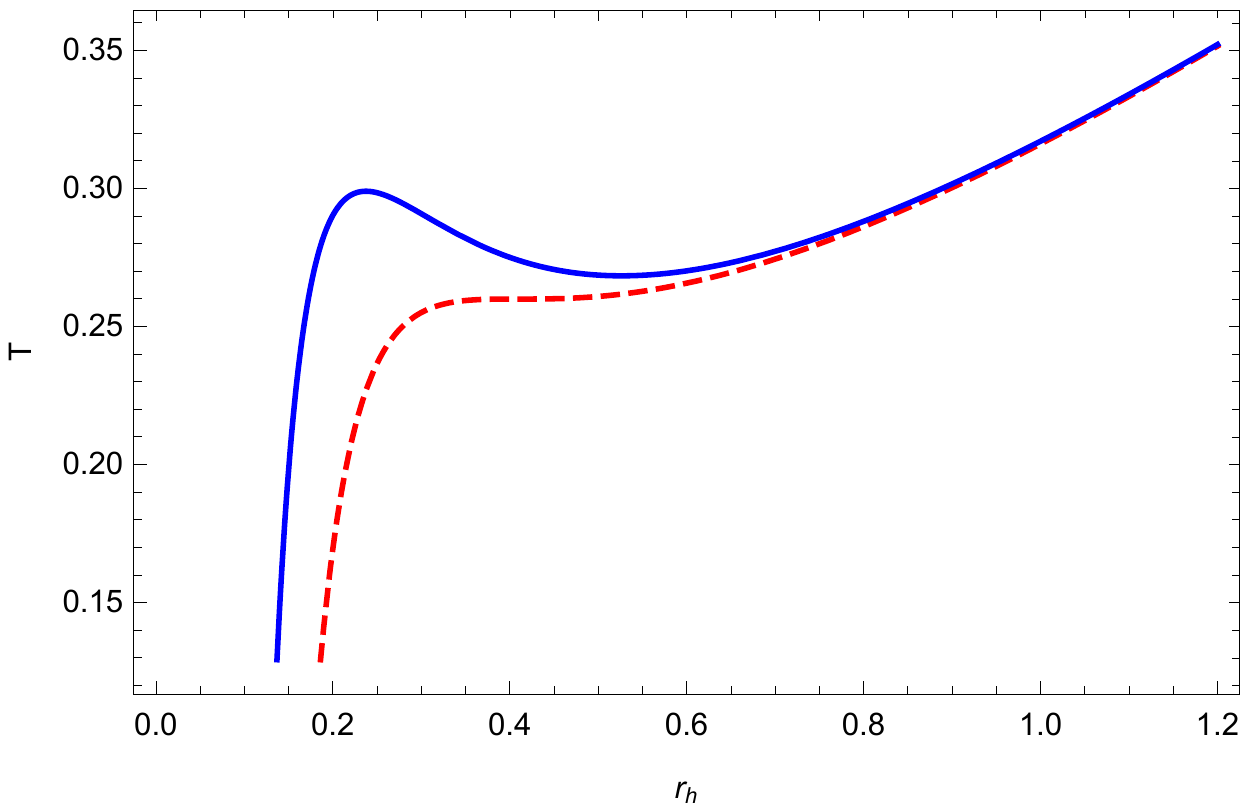}\hspace{0.3cm}
		 \includegraphics[width=5cm]{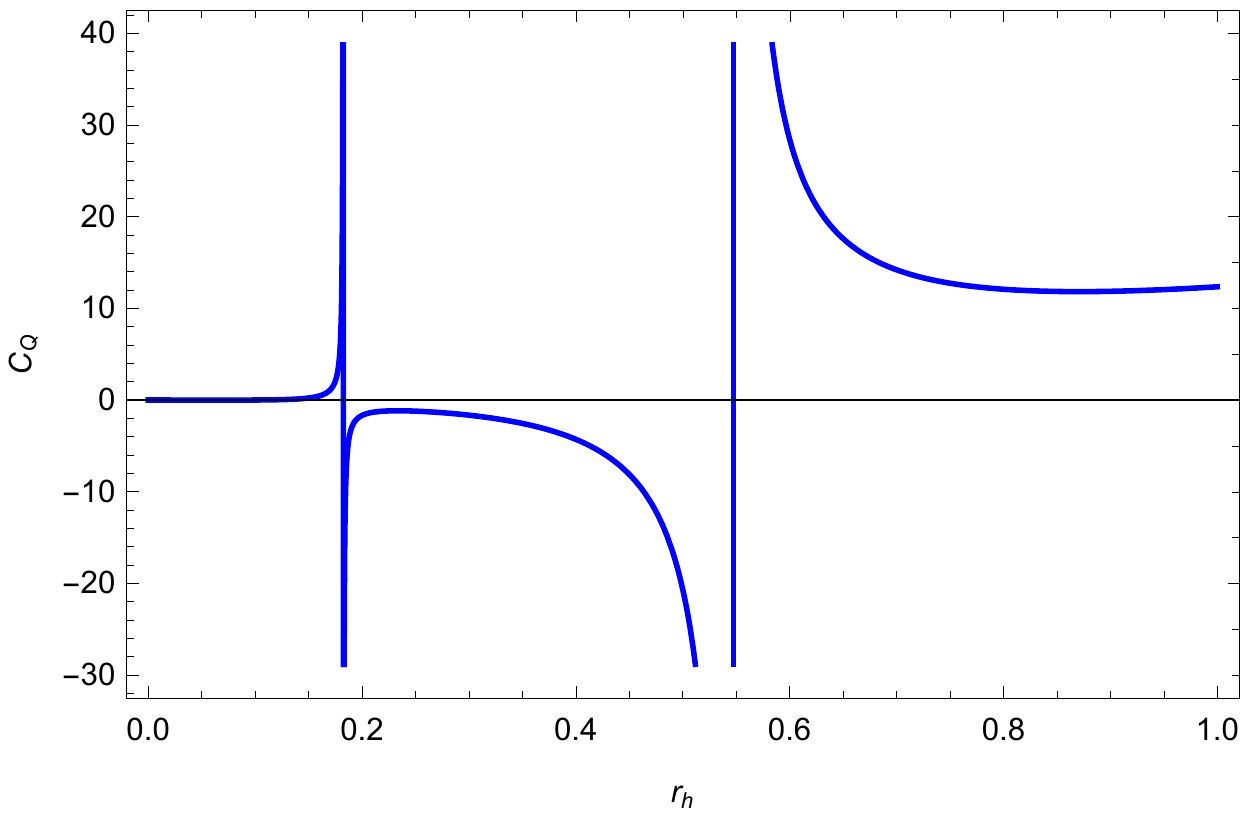}\hspace{0.3cm}
		 \includegraphics[width=5cm]{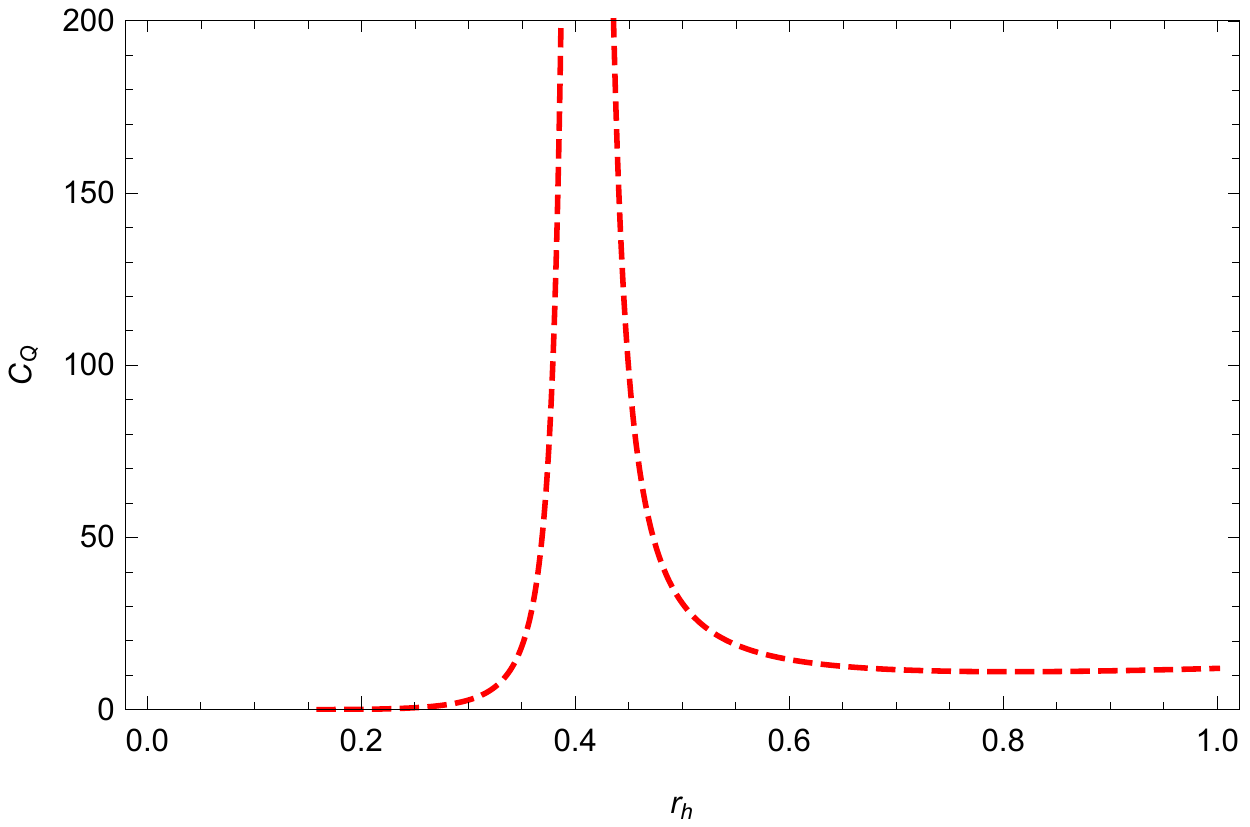}
		\caption{The temperature and heat capacity of RN-AdS black hole as the function of the horizon radius. We choose $L=1$. The blue and the red dashed curve correspond to $Q=1/8$ and $Q=1/6$, respectively. } \label{fig_TRNADS}
	}
\end{figure}

The heat capacity at constant charge is defined as
\be
C_Q=\left.\d{\partial M}{\partial T}\right|_Q=T\left.\d{\partial S}{\partial T}\right|_Q=T\left.\d{\partial S/\partial r_h}{\partial T/\partial r_h}\right|_Q=\frac{2 \pi  r_h^2 \left(L^2 r_h^2+3 r_h^4-L^2 Q^2\right)}{-L^2 r_h^2+3 r_h^4+3 L^2 Q^2}.
\ee
According to the denominator, we can derive the position of Davies points,
\be
r_h=\sqrt{\frac{L^2}{6}\pm \frac{1}{6} \sqrt{L^4-36 L^2 Q^2}}.
\ee

When $L>6Q$, there are two different roots, which correspond to the first type of Davies point. While for $L=6Q$, the two roots coincide, which is the critical point and belongs to the second type of Davies point. 

As is shown in Fig.\ref{fig_TRNADS}, in the blue curve there are two first types of Davies points. Each Davies point separates the temperature into two branches. The heat capacity has opposite signs on the two branches. As the two Davies points coincide in the red curve, it becomes a critical point. We also depict the heat capacity in this figure. At the two types of Davies point, heat capacities both diverge. According to the standard Ehrenfest classification, the phase transitions are both of second order. In the following, we will try to distinguish these phase structures using our generalized Ehrenfest classification.

\section{Phase transitions at the first type of Davies points}

According to Eq.(\ref{TRNADS}), for smaller $r_h$ the temperature of RN-AdS black hole is dominated by the electric charge $Q$ rather $L$, while for larger $r_h$ the temperature is dominated by $L$. Therefore, the behaviors of the temperature near the left Davies point are different from that near the right Davies point. We should analyze the phase structures at the two Davies points, respectively.

For simplicity, we set $L=1$ in this section. 

The left Davies point lies at
\be
T_c=-\frac{\sqrt{\frac{3}{2}} \left(12 Q^2+\sqrt{1-36 Q^2}-1\right)}{\pi  \left(1-\sqrt{1-36 Q^2}\right)^{3/2}}, \quad r_c=\frac{\sqrt{1-\sqrt{1-36 Q^2}}}{\sqrt{6}}.
\ee

%

We define some dimensionless quantities
\be
T=T_c(1+t), \quad r_h=r_c(1+\rho). \label{dimlessQ}
\ee
Now the Davies point lies at $t=\rho=0$. In this way, we can do a series expansion near the Davies point. Because this left Davies point corresponds to the maximal value of temperature, there is always $t\leq 0$.

After setting $Q=1/8$, we can obtain a dimensionless equation of state from Eq.(\ref{TRNADS}),
\be\label{Leos}
0.057 \rho ^4-\rho ^3 (0.302 t+0.0729)-\rho ^2 (0.906 t+0.224)-0.906 \rho  t-0.302 t=0.
\ee

Near the Davies point, we solve Eq.(\ref{Leos}) to obtain a series solution of $\rho$ as functions of $t$,
\bea\label{Lrho_t}
\rho&=&-1.162 (-t)^{1/2}-1.804 t -3.270 (- t)^{3/2}+6.633 t^2 -14.616 (-t)^{5/2}   +O(t^3), \no \\
\rho&=&-1.162 (-t)^{1/2}-1.804 t +3.270 (- t)^{3/2}+6.633 t^2 +14.616 (-t)^{5/2} +O(t^3).
\eea
The two series solutions correspond to the two branches on both sides of the Davies point. The most notable feature of this result is that it contains terms with fractional exponents.

According to the definition above, we can obtain free energy,
\be
F=\frac{(\rho +1)^2 \left(\rho ^2+2 \rho -2\right) \left(\sqrt{1-36 Q^2}-1\right)+18 \left(\rho ^4+4 \rho ^3+6 \rho ^2+4 \rho +4\right) Q^2}{12 (\rho +1) \sqrt{6-6 \sqrt{1-36 Q^2}}},
\ee
where $\rho=\rho(t)$ is given by Eq.(\ref{Lrho_t}).


Substituting the results in Eq.(\ref{Lrho_t}) into the free energy, we have
\bea
F&=& 0.105 -0.053 t -0.082 (-t)^{3/2} + 0.13 t^2 -0.227 (-t)^{5/2} +O(t^3), \\
F&=& 0.105 -0.053 t +0.082 (-t)^{3/2} + 0.13 t^2 +0.227 (-t)^{5/2} +O(t^3).
\eea
One should note that there are also terms with fractional exponents. Especially, the coefficients before the term $(-t)^{3/2}$ have opposite signs. 

Now it is time to calculate the derivatives of the free energy. We will use Caputo's definition of fractional derivatives\cite{Caputo.1967, Kilbas.1993, Gorenflo.1997}. The main formulae we may need are
\be
D^\alpha_x x^n =0, ~~~ \text{with $n$ integer and $\alpha>n$}; \quad D^\alpha_x x^a \propto x^{a-\alpha}, ~~~ \text{with  $a$ real}.
\ee
For more properties of fractional derivatives one can refer to the appendix of\cite{Ma.490495.2019}.

The $\alpha$-order fractional derivative of the free energy with $1<\alpha \leq 2$ have the form of
\be\label{RNDF}
D_{ t}^{\alpha}F(t)=(-t)^{1-\alpha}\left[A(-t)^{1/2}\pm B t+C(-t)^{3/2} \pm D t^2+\cdots \right],
\ee
where $A,B,C,D$ are some constant coefficients. The "+" sign corresponds to the case of approaching the Davies point from the left-hand side and the "-" sign corresponds to the other side. 

Due to $t\leq 0$, the $D_{ t}^{\alpha}F-t$ curves all turn up on the left hand-side of $t=0$ axis. One can also express $D_{ t}^{\alpha}F$ as functions of $\rho$. In this case, $D_{ t}^{\alpha}F$ can occur on both sides of $\rho=0$ axis.

\begin{figure}
	\centering{
		\includegraphics[width=7cm]{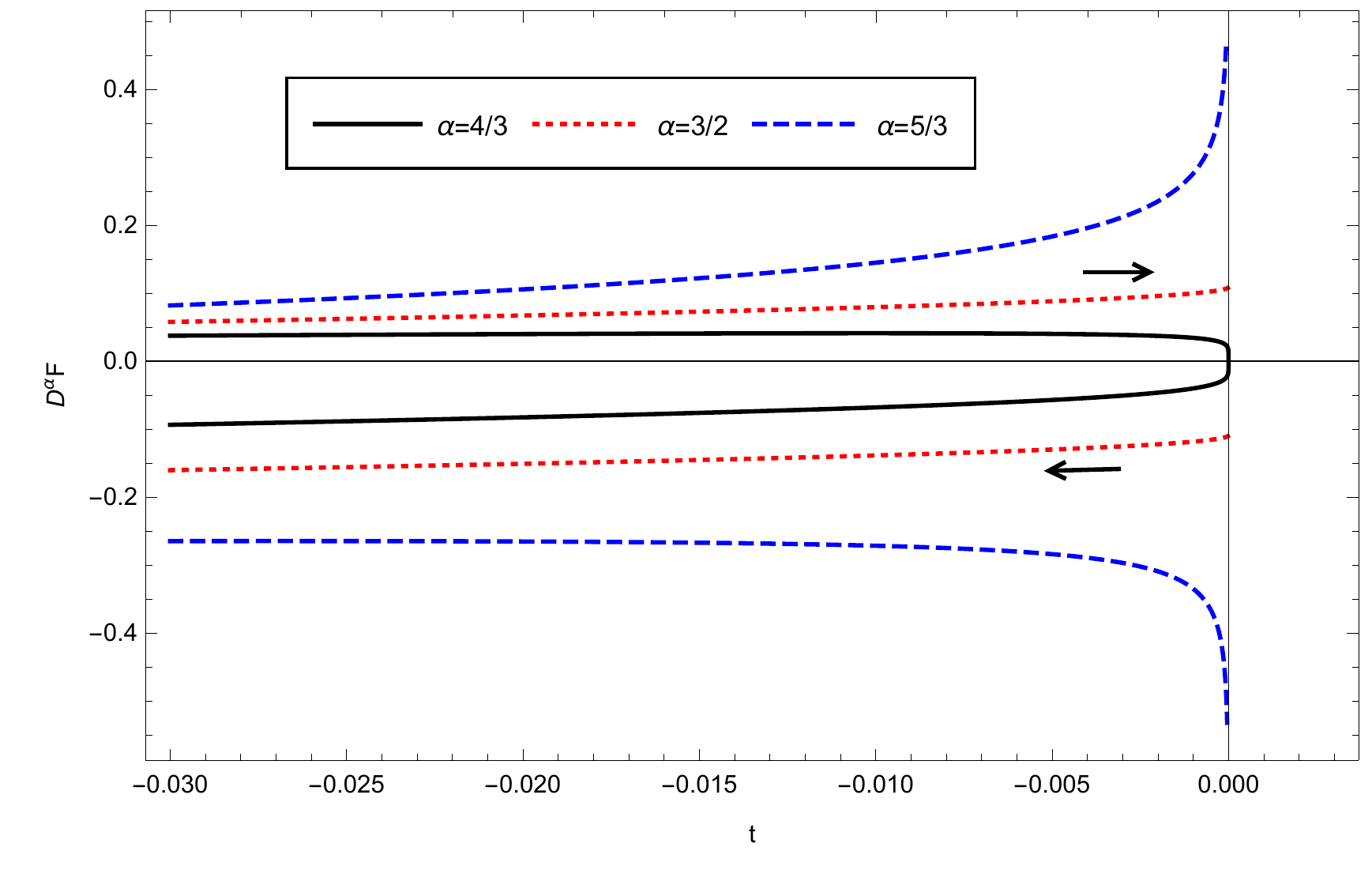}  \hspace{0.5cm}
		\includegraphics[width=7cm]{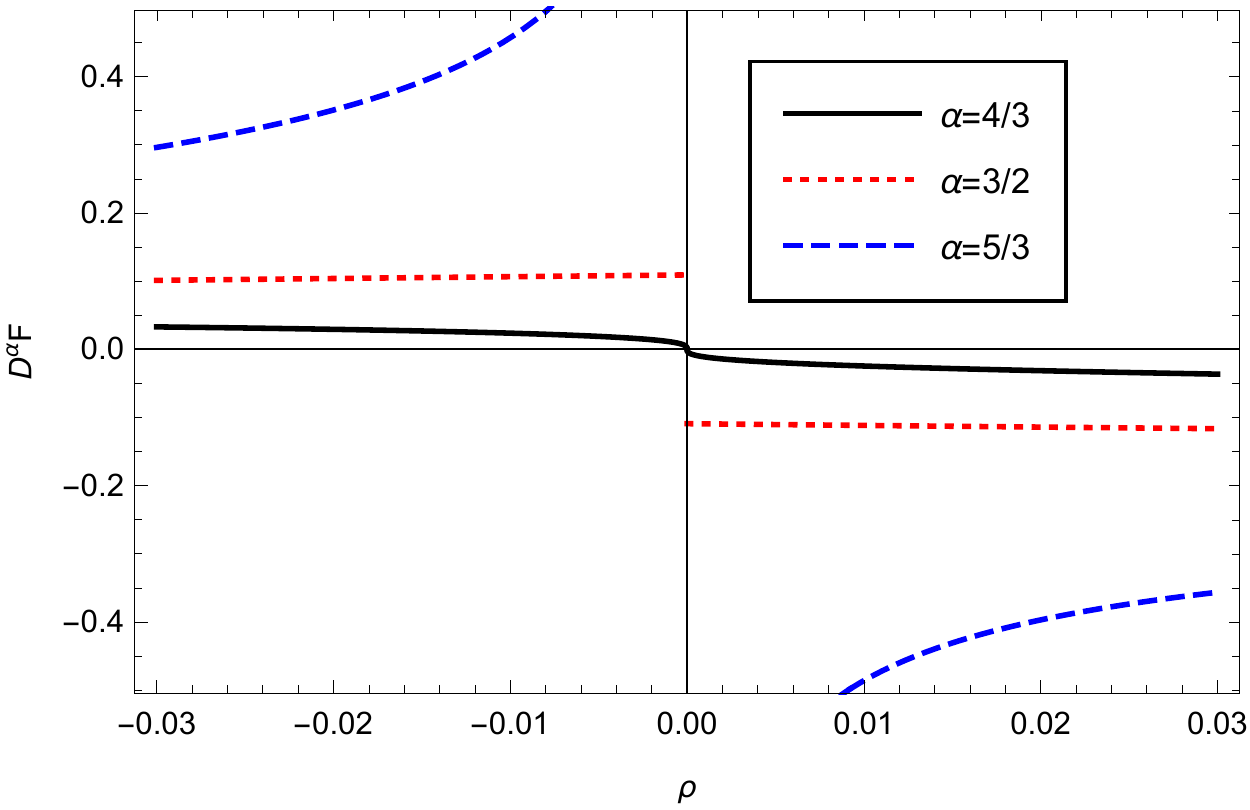} 
		\caption{The behaviors of $D_{ t}^{\alpha}F(t)$ near the left Davies point for RN-AdS black hole. The left panel depicts the $D_{ t}^{\alpha}F-t$ curves. The black arrows indicate the direction of increasing $r_h$. The right panel displays the $D_{ t}^{\alpha}F-\rho$ curves.} \label{fig_dftL}
	}
\end{figure}

Substituting concrete values of $\alpha$ into Eq.(\ref{RNDF}) and taking the $t\rightarrow 0^{-}$ limit, we can obtain
\be
\lim_{ t\rightarrow 0^{-}}D_{ t}^{\alpha}F(t)=\left\{
\begin{array}{lr}
	0 ~~&\text{for}~~\alpha<3/2,\\
	\pm 0.291 ~~&\text{for}~~\alpha=3/2,\\
	\pm \infty ~~ &\text{for}~~\alpha>3/2.
\end{array}
\right.
\ee

A look at the behaviors of $D_{ t}^{\alpha}F(t)$ in Fig.\ref{fig_dftL} shows that when $\alpha<3/2$ the $D_{ t}^{\alpha}F(t)$ is continuous at the Davies point, when $\alpha=3/2$ the $D_{ t}^{\alpha}F(t)$ has a finite jump  and when $\alpha>3/2$ the $D_{ t}^{\alpha}F(t)$ is divergent. According to the generalized classification, a $3/2$-order phase transition occurs at this Davies point.

Next, we analyze the phase structure of the RN-AdS black hole at the right Davies point. The procedure is similar. In this case, Davies point turns up at 
\be
T_c=\frac{\sqrt{\frac{3}{2}} \left(1-12 Q^2+\sqrt{1-36 Q^2}\right)}{\pi  \left(1+\sqrt{1-36 Q^2}\right)^{3/2}}, \quad r_c=\frac{\sqrt{1+\sqrt{1-36 Q^2}}}{\sqrt{6}}.
\ee

With the dimensionless quantities, the equation of state is reduced to
\be\label{Reos}
1.380 \rho ^4+(2.573-2.948 t)\rho ^3 +(1.099 -8.844 t)\rho ^2-8.844 \rho  t-2.948 t=0.
\ee
It should be noted that the right Davies point of RN-AdS black hole corresponds to the minimal value of temperature, thus we always have $t\geq 0$. 


In the same way, we solve Eq.(\ref{Reos}) to obtain $\rho=\rho(t)$ in series form and substitute them into the free energy. At last, we can derive the free energy in the following form
\bea
F&=& 0.117-0.233 t-0.510 t^{3/2}-0.519 t^2-0.397 t^{5/2}   +O(t^3), \\
F&=& 0.117-0.233 t+0.510 t^{3/2}-0.519 t^2+0.397 t^{5/2} +O(t^3).
\eea

\begin{figure}
	\centering{
		\includegraphics[width=7cm]{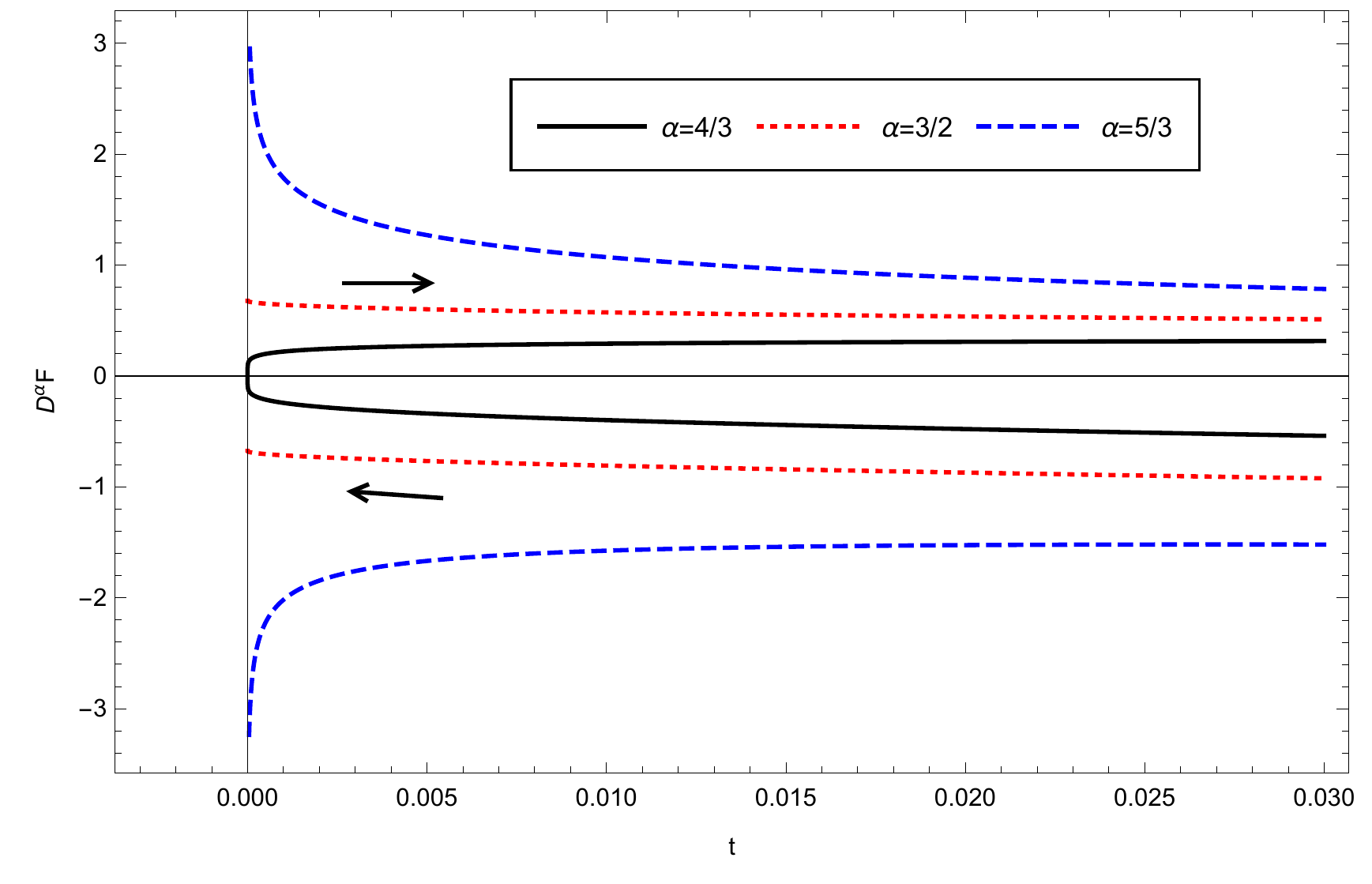} \hspace{0.5cm}
		\includegraphics[width=7cm]{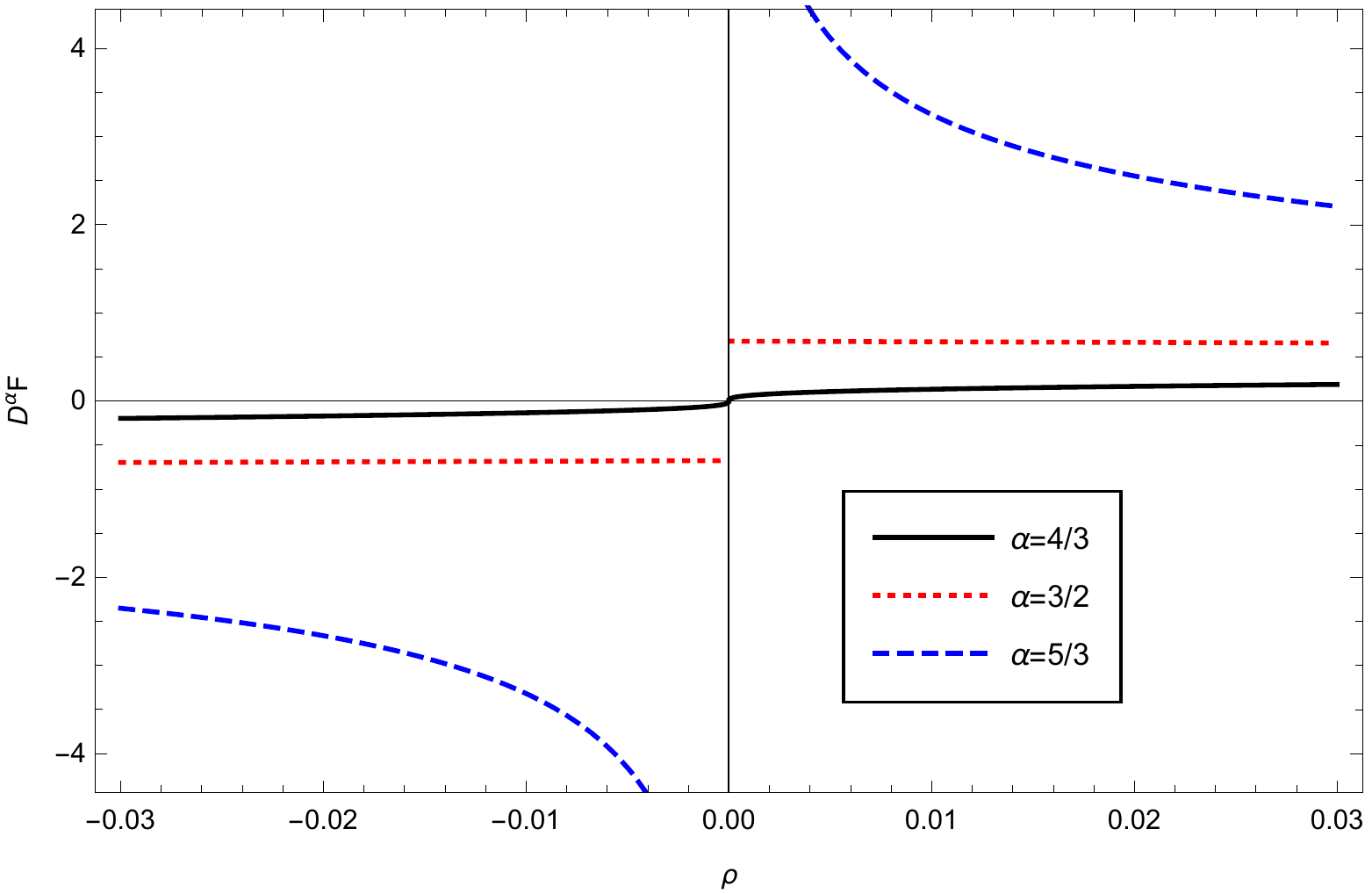} 
		\caption{The behaviors of $D_{ t}^{\alpha}F(t)$ near the right Davies point for RN-AdS black hole. The left panel depicts the $D_{ t}^{\alpha}F(t)-t$ curves. The black arrows indicate the direction of increasing $r_h$. The right panel displays the $D_{ t}^{\alpha}F(t)-\rho$ curves.} \label{fig_dftR}
	}
\end{figure}

Except for the values of the coefficients and $t\geq 0$, these free energies have similar forms to those near the left Davies point. Therefore, their fractional derivatives are also analogous. We do not write out them here. In Fig.\ref{fig_dftR}, we directly depict the behaviors of the fractional derivatives of free energy as functions of $t$ or $\rho$. The $t \rightarrow 0^{+}$limit values of the fractional derivatives are
\be
\lim_{ t\rightarrow 0^{+}}D_{ t}^{\alpha}F(t)=\left\{
\begin{array}{lr}
	0 ~~&\text{for}~~\alpha<3/2,\\
	\pm 0.678 ~~&\text{for}~~\alpha=3/2,\\
	\pm \infty ~~ &\text{for}~~\alpha>3/2.
\end{array}
\right.
\ee

In summary, although the temperatures have different behaviors at the left and right Davies points, the phase structures at the two points are similar. However, phase transitions at these points are both $3/2$-order, but not second-order as people thought before. To our knowledge, in usual thermodynamic systems temperatures are generally monotonic and have no analogues of the first type of Davies point.

\section{Phase transitions at the second type of Davies point}

Now we consider the second type of Davies point, namely the critical point, where 
\be
dT/dr_h=0, \quad d^2T/dr_h^2=0.
\ee
The critical point lies at
\be
r_c=L/\sqrt{6}, \quad Q_c=L/6, \quad T_c=\sqrt{\d{2}{3}}\d{1}{\pi L}.
\ee
We also use the dimensionless quantities defined above. Because the critical point is an infection point, now the dimensionless temperature $t$ can be greater or smaller than zero. 

 On the basis of the definition of Eq.(\ref{dimlessQ}), we also define $Q=Q_c(1+q)$. The equation of state is simplified to
\be\label{eosRNADS}
2q+q^2+8t+24 t \rho+24 t\rho ^2+(8t-4)\rho^3-3 \rho ^4=0.
\ee
It is a quartic equation about $\rho$. In principle, we can analytically solve it. But the exact solutions are too lengthy and unnecessary.

Near the critical point, the free energy has the form
\be
F=\frac{L \left(-\rho ^4-4 \rho ^3+3 q^2+8 \rho +6 q+8\right)}{24 \sqrt{6} (\rho +1)}.
\ee
For simplicity, we introduce a reduced free energy
\be\label{redF}
\tilde{F}=\d{24\sqrt{6}}{L}F=\frac{ -\rho ^4-4 \rho ^3+3 q^2+8 \rho +6 q+8}{ \rho +1}.
\ee

Next, we may have two approaches for further calculations. The first one is to take the same procedure as the former section, first solve the equation of state to obtain $\rho=\rho(t,q)$ and then substitute it into Eq.(\ref{redF}) to obtain the free energy as functions of $(t,q)$. But here we choose another method used by Hilfer\cite{Hilfer.2000}. In the $(t,q)$ plane, there are infinitely many possible ways of approaching the critical point. We can reparametrize the dimensionless quantities,  
\be\label{para}
t=x \cos\theta, \quad q=x \sin\theta,
\ee
where $\theta$ controls the direction of approaching the critical point, and $x \rightarrow 0$ leads to $t \rightarrow 0$ and $q \rightarrow 0$.

Correspondingly, in this new coordinate $(x,\theta)$, we have
\be
\tilde{F}=\tilde{F}(t,q)=\hat{F}(x,\theta).
\ee

In this way, we can solve Eq.(\ref{eosRNADS}) to obtain 
\bea\label{rhoxth}
\rho&=&\rho(x,\theta)=\frac{x^{1/3} (\sin \theta+4 \cos \theta)^{1/3}}{2^{1/3}}+\frac{x}{32} (3 \sin \theta+28 \cos \theta)+\frac{x^{2/3} (12 \cos \theta-\sin \theta)}{4\times 2^{2/3} (\sin \theta+4 \cos \theta)^{1/3}} \no \\
&+&\frac{x^{4/3} \left(29 \sin ^3\theta+2368 \cos ^3\theta+1136 \sin \theta \cos ^2\theta-4 \sin ^2\theta \cos \theta\right)}{384 \times 2^{1/3} (\sin \theta+4 \cos \theta)^{5/3}}+O(x^{5/3}).
\eea
But it should be noted that this series solution is valid for $\sin \theta+4 \cos \theta \neq 0$, namely $q+4t \neq 0$. 
Substituting Eq.(\ref{para}) and Eq.(\ref{rhoxth}) into Eq.(\ref{redF}), we obtain
\bea\label{Fxth}
\hat{F}(x,\theta)&=&8+(4 \sin \theta -8 \cos \theta)x-\frac{3  (\sin \theta +4 \cos \theta )^{4/3}x^{4/3}}{2^{1/3}} \no \\
&+&\frac{3  (\sin \theta -4 \cos \theta ) (\sin \theta +4 \cos \theta )^{2/3}x^{5/3}}{2^{2/3}}+\frac{[8 \sin (2 \theta )-309 \cos (2 \theta )-299]x^2}{32}\no \\
&+& O(x^{7/3}).
\eea

\begin{figure}
	\centering{
		\includegraphics[width=7cm]{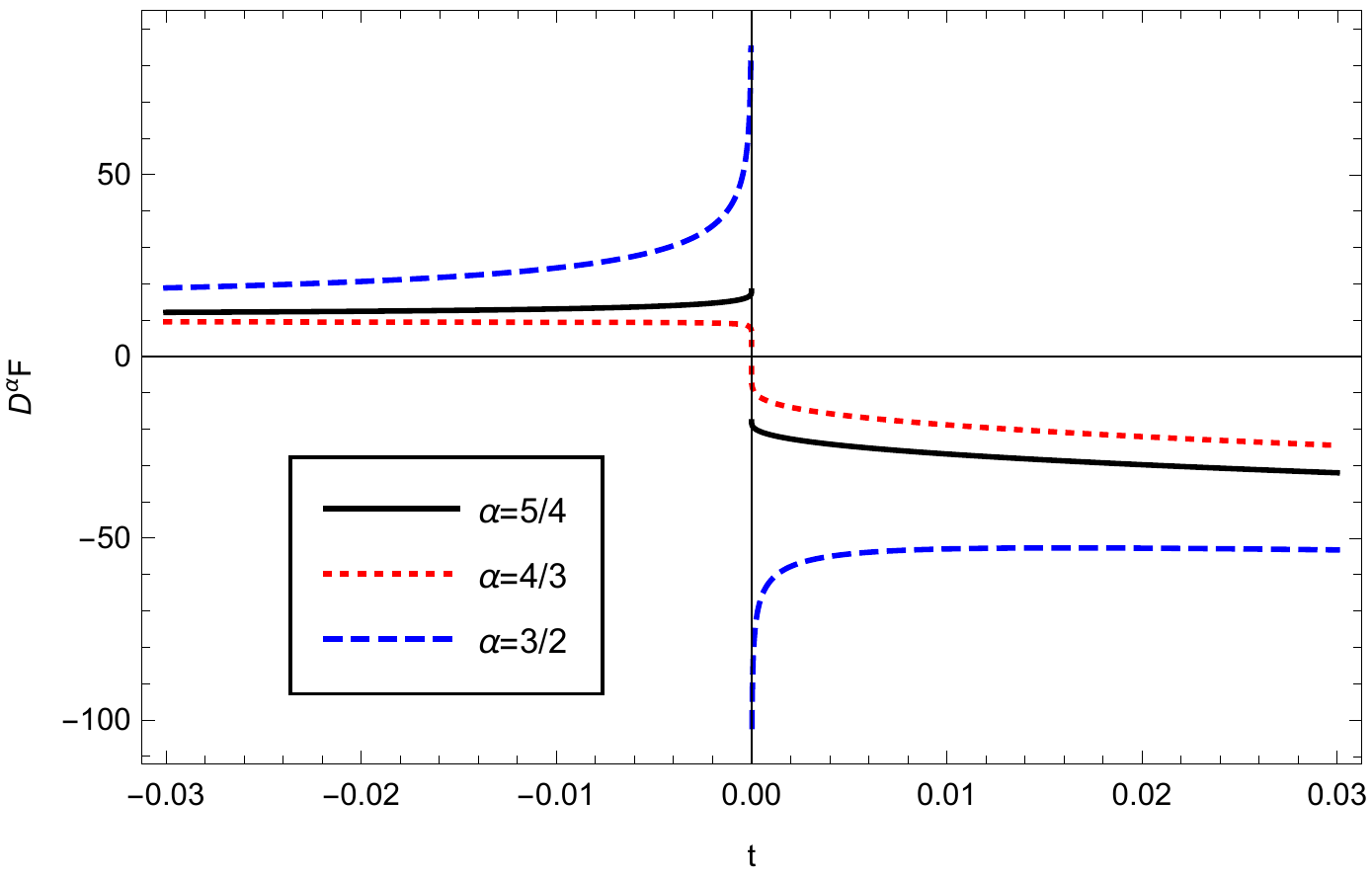} 
		\caption{The behaviors of $D_{ t}^{\alpha}\tilde{F}(t)$ near the critical point for RN-AdS black hole. } \label{fig_FtRNADS}
	}
\end{figure}

Now we can calculate the fractional derivatives of the free energy and take the $x\rightarrow 0^{+}$ limit. The results are as follows,
\be
\lim_{ x\rightarrow 0^{+}}D_{ x}^{\alpha}\hat{F}(x,\theta)=\left\{
\begin{array}{lr}
	0 ~~&\text{for}~~\alpha<4/3,\\
	-\frac{4\times 2^{2/3} \pi  (\sin \theta +4 \cos \theta )^{4/3}}{3 \sqrt{3} \Gamma \left(\frac{2}{3}\right)} ~~&\text{for}~~\alpha=4/3,\\
	\pm \infty ~~ &\text{for}~~\alpha>4/3,
\end{array}
\right.
\ee
Except for those values of $\theta$ satisfying $\sin \theta+4 \cos \theta = 0$, the $4/3$-order derivative of the free energy is discontinuous, which is a signature of $4/3$-order phase transition. For clarity, we can choose a concrete direction to depict the fractional derivative. 
We  take $\theta=0$ and $\theta=\pi$, which correspond to the $t$-axis, thus $q=0$. Substituting them into Eq.(\ref{Fxth}), we can obtain
\be
\tilde{F}(t)=\left\{
\begin{array}{lr}
	8-8t-12\times 2^{1/3} t^{4/3}-12\times 2^{2/3} t^{5/3}-19 t^2 +O(t^{7/3}), ~~& ~~t>0,\\
	8-8t-12\times 2^{1/3} (-t)^{4/3}+12\times 2^{2/3} (-t)^{5/3}-19 t^2 +O(t^{7/3}), ~~& ~~t<0.
\end{array}
\right.
\ee
Due to the $(\pm t)^{4/3}$ terms, there is indeed a finite jump for $\alpha=4/3$,
\be
\lim_{t\rightarrow 0^{-}}\d{d^\alpha \tilde{F}}{dt^\alpha}=\frac{32 \sqrt[3]{2} \pi }{3 \sqrt{3} \Gamma(2/3)} \neq\lim_{t\rightarrow 0^{+}}\d{d^\alpha \tilde{F}}{dt^\alpha}=-\frac{32 \sqrt[3]{2} \pi }{3 \sqrt{3} \Gamma(2/3)}.
\ee
According to the generalized classification, the phase transition at the critical point is of $4/3$ order.  The result is exhibited in Fig.\ref{fig_FtRNADS}.

Now we need to deal with the exceptional case, $t+4q=0$, which selects a special direction in the $(t,q)$ plane. We need to approach the critical point along this direction from both sides. 

In this case, the equation of state becomes
\be
-3 \rho ^4+\rho ^3 (8 t-4)+24 \rho ^2 t+24 \rho  t+16 t^2=0.
\ee
Its series solution is 
\be
\rho(t)=-\frac{2 t}{3}-\frac{40 t^2}{81}-\frac{470 t^3}{729}+O(t^4).
\ee
One can see that in this special case there are no terms with fractional exponents.  It naturally leads to a free energy without fractional exponents
\be
\tilde{F}(t)=8-24t+32 t^2+\frac{32 t^3}{3}+O(t^4).
\ee

Clearly, in the $t\rightarrow 0$ limit,  the arbitrary fractional order derivatives of this free energy is zero and integral order derivatives are constants, thus the free energy is always continuous. This exceptional direction is called phase boundary at the critical point\cite{Hilfer.2000}, but where it is a second-order phase transition along this phase boundary.

%

\section{Discussions and Conclusions}
\label{Conclusions}

Based on the generalized Ehrenfest classification of phase transition, we analyzed the phase structure of RN-AdS black hole at the Davies points.
There should be two types of Davies points, corresponding to the extreme points and infection points of the temperature, respectively. According to standard Ehrenfest classification, it is second-order phase transitions occurring at these points. However, by means of fractional derivatives we found that at the first type of Davies points, the phase transition is of $3/2$ order. At the second types of Davies point, it is a $4/3$-order phase transition, which is the same as the result we obtained in extended phase space\cite{Ma.490495.2019}. Therefore, this generalized classification can indeed reflect more details on the critical behaviors of black holes.  

RN black hole and Schwarzschild-AdS(SAdS) black hole can be taken as the $\Lambda \rightarrow 0$ and $Q \rightarrow 0$ limits of RN-AdS black hole, respectively. The two black holes both have the first type of Davies point. We also studied the fractional phase transition of these black holes and found that they are both of $3/2$ order. Indeed, if we divide the temperature curve of the RN-AdS black hole into two parts, the left part has similar behaviors to that of the RN black hole and the right part is similar to the temperature of the SAdS. So it is not surprising that they have the same order of phase transition. 

We used to think that the equations of state determine the order of fractional phase transition. But, according to \cite{Chabab.620430.2021} it seems that what influences the order of phase transition is the symmetry of black hole spacetimes. In the present work, one can see that the equations of state near the first and second type of Davies points are both quartic equations.  To clarify this problem, we need to further consider other black holes.  Although for RN-AdS black hole it is a $3/2$-order phase transition at the first type of Davies point, we do not know whether the order is the same for other black holes with the first type of Davies point.  Until now, the generalized Ehrenfest classification is only used to deal with phase transitions at Davies points of black holes. It should also be available for reanalyzing the original first-order phase transition. The orders of phase transitions at critical points are related to critical exponents. In real physical system, these exponents can be explored in laboratory. The critical exponents corresponding to the fractional order phase transitions of black holes should also be examined.  These interesting problems should be considered in future works.

\bigskip
\bigskip



\acknowledgments
We are grateful to the reviewers for their suggestions that helped to improve the content of the paper. This work is supported in part by Shanxi Provincial Natural Science Foundation of China (Grant No. 201701D121002), by Scientific and Technological Innovation Programs of Higher Education Institutions in Shanxi (Grant No. 2021L386) and by Datong City Key Project of Research and Development of Industry of China (Grant No. 2018021). 

\bibliographystyle{JHEP}
\bibliography{E:/mms/References/phase_transition}

\end{document}